\documentclass{elsart}

\usepackage{graphicx}

\begin{document}

\begin{frontmatter}


% use the thanksref command within \title, \author or \address for footnotes:
% \author{Name\thanksref{label2}}
% including your email address:
% \address{Address\thanksref{email}}
% \thanks[email]{E-mail: }

\title{Electron Transport in Hybrid Ferromagnetic/Superconducting Nanostructures}


\author{V. T. Petrashov, I. A. Sosnin, C. Troadec, I. Cox, and A. Parsons}

\address{Department of Physics, Royal Holloway, University of London, Egham, Surrey TW20 0EX, U. K.}

\begin{abstract}
\par We observe large amplitude changes in the resistance of ferromagnetic ($F$) wires
at the onset of superconductivity of adjacent superconductors ($S$). New sharp peaks of
large amplitude are found in the magnetoresistance of the $F$-wires. We discuss a new
mechanism for the long-range superconducting proximity effect in $F/S$ nanostructures
based on the analysis of the topologies of actual Fermi-surfaces in ferromagnetic metals.
\end{abstract}

\end{frontmatter}
\section{Introduction}
\par Among the highlights of recent investigations in the field of mesoscopic
superconductivity is the discovery of the proximity effects in disordered ferromagnetic
(F) conductors \cite{1,2,3,4,5,6}. Several theoretical works have been published recently
\cite{7,8,9}, giving an account for some of the features of the observed effects, however
the origin of the large decrease in the resistance of ferromagnetic wires \cite{4} and
long range effects \cite{1} are yet to be explained. In this paper we report new results
on the dependence of the proximity effects in diffusive mesoscopic $F/S$ structures on
applied magnetic. To separate the contribution of the bulk and interface effects we
measure simultaneously the barrier resistance of the $F/S$ interfaces and the resistance
of the $F$-wires. The amplitude and character of the observed effects cannot be explained
on the base of the interface phenomena, suggested recently in \cite{10}. We discuss a new
mechanism for the long-range superconducting proximity effect in
ferromagnetic/superconducting nanostructures based on the analysis of the topologies of
actual Fermi-surfaces in ferromagnetic metals.

\section{Experimental}

\par The geometry of the structures is shown in Fig. 1. The structures were made by
multi-layer e-beam lithography. The first layer was Ni wire in contact with Al wires of
the second layer. We measured the resistance of Ni wires using pairs of Al wires as
potential leads with current flowing through Ni wire from point $a$ to $b$, in the
temperature range from 0.28K to 6K in magnetic fields up to 5T applied perpendicular to
the substrate. The value of  r for the Ni and Al films was about 12 $\mu\Omega$cm and 0.8
 $\mu\Omega$cm, and diffusion constants, $D$, of about 40 cm$^2$/s and 200 cm$^2$/s,
 correspondingly.

\begin{figure}[h]
%h=here, t=top, b=bottom, p=separate figure page
\begin{center}\leavevmode
\includegraphics[width=0.8\linewidth]{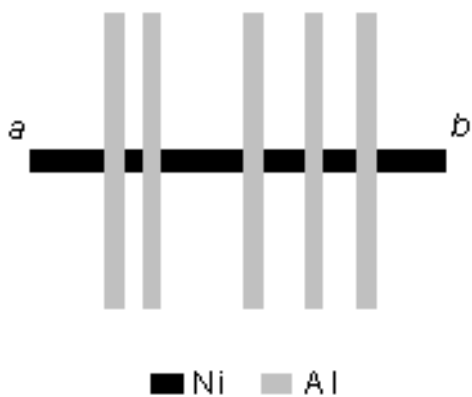}
\caption{Sample geometry.} \label{Fig. 1.}
\end{center}
\end{figure}

\par The dependence of the resistance of one of our Ni wires on temperature is shown in
Fig. 2a. The sheet resistance of Ni and Al wire was 5 $\Omega$ and 0.15 $\Omega$,
respectively. The width, $w$ and thickness, $t$ of Ni wire were 250 nm and 22 nm. The
width and thickness of Al potential leads were 100 nm and 55 nm with distance, $L$, of
1000 nm between them. It is seen from Fig. 2a that the value of the resistance of the Ni
wire in the minimum is lower than that in normal state by 1.2 $\Omega$. The total
peak-to-peak amplitude of the effect is more than 2 $\Omega$. The changes in the
resistance are much larger that the Ni/Al barrier resistance in both the normal and
superconducting state (see Fig. 2c). The sign of the effect for the wire is opposite to
that for both barriers involved: the resistance of the barrier in superconducting state
is higher than that in normal, while the resistance of the wire is lower. An example of
the dependence of the zero bias resistance for the same sample on magnetic field is shown
in Fig. 2b. One of the striking features of the dependence is the sharp peak at 110 G.
\begin{figure}[h]
%h=here, t=top, b=bottom, p=separate figure page
\begin{center}\leavevmode
\includegraphics[width=0.8\linewidth]{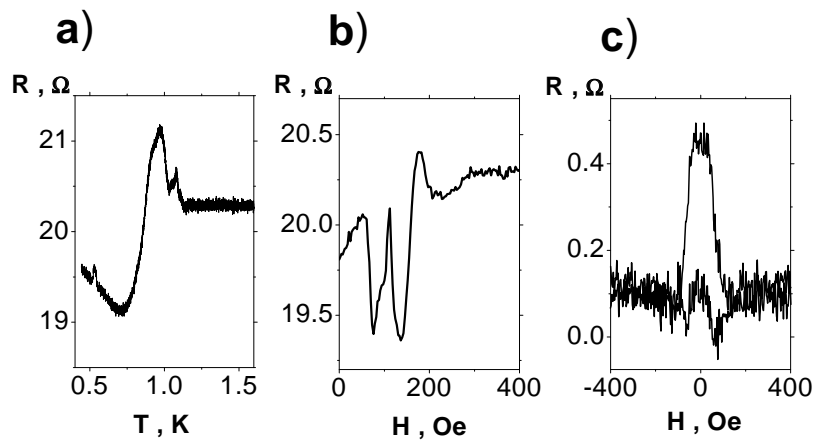}
\caption{(a) Temperature dependence of the resistance of one of the samples in zero
magnetic field. Ni wire parameters are $L$=1000nm, $w$=250nm, $t$=22nm. (b)
Magnetoresistance of the same sample at $T$=0.27K. (c) Magnetoresistance of the two
barriers involved.} \label{Fig. 2.}
\end{center}
\end{figure}
\begin{figure}[h]
%h=here, t=top, b=bottom, p=separate figure page
\begin{center}\leavevmode
\includegraphics[width=0.8\linewidth]{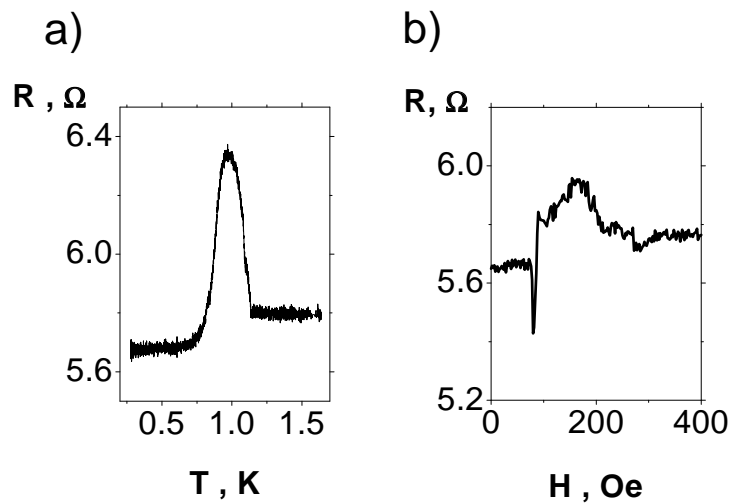}
\caption{(a) Temperature dependence of the resistance of another sample in zero magnetic
field. Ni wire parameters are $L$=400nm, $w$=250nm, $t$=40nm. (b) Magnetoresistance of
the same sample at $T$=0.27K.} \label{Fig. 3.}
\end{center}
\end{figure}
\par Figures 3a and 3b show the
dependence of the resistance on magnetic field and temperature of the segment of Ni wire
250 nm wide 40 nm thick with distance between the potential leads of 400 nm long. The
sheet resistance of Ni and Al were 3.4 $\Omega$ and 0.15 $\Omega$, respectively. A sharp
dip is seen at the magnetic field of 70 G. The dependence of the resistance on
temperature (Fig. 3a) shows maximum followed by a plateau. The low-temperature value of
the resistance is lower than that in the normal state. We observe sharp features in the
differential resistance $dV/dI$ vs current, $I$, at the magnetic field of 70 G (Fig. 4).
Figs. 5a and 5b show the temperature dependence of the resistance of the barriers,
$R_{b}$, between Ni and Al wires and their voltage-current characteristics. For this
sample the value of $R_{b}$ increases from 0.05 $\Omega$ by more that order in magnitude
when Al wire goes superconducting. Note, that the temperature dependence of the barrier
resistance may strongly deviate from the dependence of the model contacts suggested in
\cite{10}, described by an admixture of a tunnel and ballistic contact. While the
temperature dependencies of the contacts shown in Fig. 5 follow qualitatively the "mixed"
model at temperatures close to $T_{c}$, at lowest temperatures an unexpected drop in the
resistance may take place (see upper curve).

\begin{figure}[h]
%h=here, t=top, b=bottom, p=separate figure page
\begin{center}\leavevmode
\includegraphics[width=0.8\linewidth]{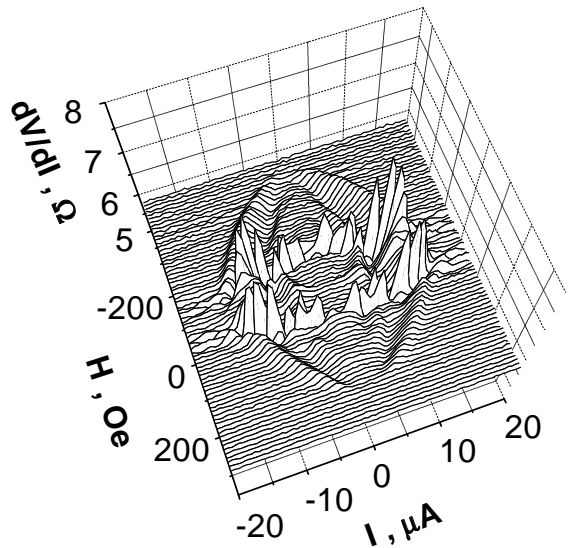}
\caption{Differential resistance, $dV/dI$, of the Ni wire of Fig. 3 as a function of
applied current and magnetic field at $T$=0.27K.} \label{Fig. 4.}
\end{center}
\end{figure}
\begin{figure}[h]
%h=here, t=top, b=bottom, p=separate figure page
\begin{center}\leavevmode
\includegraphics[width=0.8\linewidth]{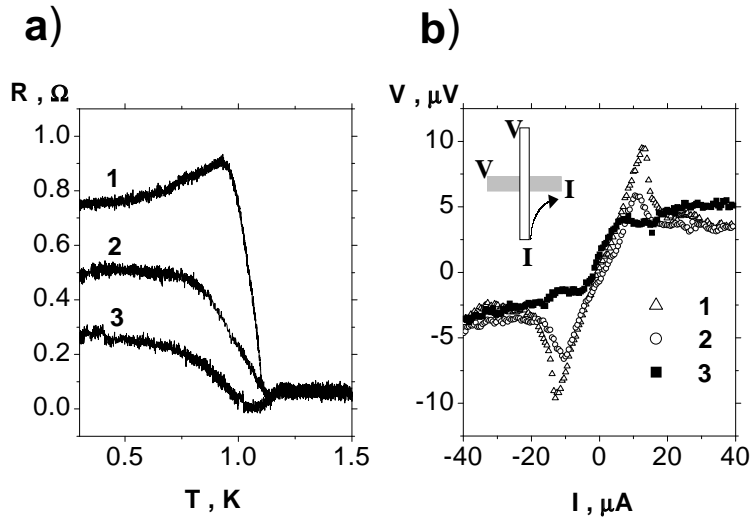}
\caption{(a) Temperature dependencies of the three barriers for the sample of Fig. 3.
Barriers 1 and 2 belong to the structure of Fig. 3. (b) Voltage-current characteristics
of the same barriers.} \label{Fig. 5.}
\end{center}
\end{figure}
\par It is seen from Figures 2 and 3 that
the behaviour of the resistance of $F$-wires below the critical temperature is strongly
sample-dependent. We find that the amplitude of the changes and even their sign may
differ from sample to sample, however, the main features, including the absence of
correlation between the resistance of the barriers and that of the wires are common for
all ten samples we have measured. A comparison of the dependence on temperature and
magnetic field of the resistance of barriers (Fig. 2c and 5) and that of the
ferromagnetic wires (Fig. 2a, b and 3) shows that the effect in wires cannot be accounted
for by the changes in the barrier resistance alone as suggested in \cite{10}.

\section{Discussion}
\par The observed effects can be explained by the changes in the
bulk resistance of the ferromagnetic wires. An additional confirmation of that comes from
the measurements with Ni wires of different thickness at the same thickness of Al wires.
We find that the peak-to-peak resistance change on average increases with the decrease in
the Ni wire thickness.
\par The theories developed so far are based on several fundamental differences between the
$N/S$ and $F/S$ systems. Only a fraction of electrons can be Andreev reflected at an
$F/S$ interface making it an effective spin filter. Furthermore, in a ferromagnet with
the exchange field energy, $h_{0}$ , the Andreev reflected quasi-particles acquire a
momentum of $Q=2h_{0}/v_{F}$ where $v_{F}$ is the Fermi velocity \cite{11}. The latter
results in an exponential decay of the superconductor-induced wave functions in diffusive
conductors over microscopic distances, $\xi_{m}=\sqrt{\hbar D/2\pi k_{B}T_{0}}$, where
$T_{0} \approx h_{0}/k_{B}$ is the Curie temperature, $D$ is the diffusion constant.
Hence the amplitude of long-range effects in diffusive $F/S$ systems with small
superconducting gap, $\Delta \ll h_{0}$ , is predicted to be negligibly small.
\par The above arguments are based on idealised isotropic models of the Fermi-surfaces of
ferromagnetic metals. However, in real metals the exchange interaction is anisotropic
with the value of $Q$ strongly depending on the position of the Andreev reflected
electrons on the Fermi-surface. We emphasise that the value of $Q$ may vanish for certain
electron pairs with opposite spins and directions of momentum \cite{12}. Moreover, there
may exist mixed-spin "pockets" of the Fermi surface \cite{13}. We argue that such groups
of electrons may be Andreev-reflected with no effects of the exchange interaction and
hence originate the long-range proximity effects in hybrid $F/S$ nanostructures.
\par The amplitude of the superconducting condensate functions on the $F$-side should depend
on the life times of the electrons in the mixed-spin groups, that in principle can be
larger than the averaged over the Fermi surface transport relaxation time (see e.g.
\cite{14} and references therein).
\par The mechanism suggested above together with the processes at the $F/S$ interfaces
\cite{8,9,10} and the dependence of the number of mixed spin electrons on magnetic field
[15] may account for the sample specific variations of the effects in our polycrystalline
wires.
\par We acknowledge financial support from the EPSRC (Grant Ref: GR/L94611).

\label{}

\end{document}